\documentclass[%
 aip,
 jcp,%
 amsmath,amssymb,
preprint,%
]{revtex4-1}
\usepackage{epsf}
\usepackage{ifthen}
\usepackage[usenames]{color}
\usepackage{amssymb,amsmath,amsbsy}
\usepackage{amsfonts}
\usepackage{multirow}
\usepackage{amsmath}
\usepackage{graphics, graphicx}
\usepackage{bm}
\usepackage{soul}
\renewcommand{\v}[1]{\ensuremath{\mathbf{#1}}} 
\renewcommand{\t}[1]{\ensuremath{\text{#1}}} 

\begin{document}
\title{Equation of Motion for the Solvent Polarization Apparent Charges in the Polarizable Continuum Model: Application to Time-Dependent CI}

\author{Silvio Pipolo}
\email{silvio.pipolo@impmc.upmc.fr}
\affiliation{Institut de Min\'{e}ralogie, de Physique des Mat\'{e}riaux et de Cosmochimie, Universit\'{e} Pierre et 
Marie Curie - Sorbonne Universit\'{e}s, Paris, France}
\author{Stefano Corni}
\email{stefano.corni@nano.cnr.it}
\affiliation{CNR-NANO Institute of Nanoscience, Modena, Italy}
\author{Roberto Cammi}
\email{roberto.cammi@unipr.it}
\affiliation{Department of Chemistry, Universit\`{a} degli studi di Parma, Parma, Italy}

\begin{abstract}
The dynamics of the electrons for a molecule in solution is coupled to the dynamics of its polarizable environment, i.e., the solvent. To theoretically investigate such electronic dynamics, we have recently developed equations of motion (EOM) for the apparent solvent polarization charges that generate the reaction field in the Polarizable Continuum Model (PCM) for solvation and we have coupled them to a real-time time-dependent density functional theory (RT TDDFT) description of the solute [Corni et al. J. Phys. Chem. A 119, 5405 (2014)]. Here we present an extension of the EOM-PCM approach to a Time-Dependent Configuration Interaction (TD CI) description of the solute dynamics, which is free from the qualitative artifacts of RT TDDFT in the adiabatic approximation. As tests of the developed approach, we investigate the solvent Debye relaxation after an electronic excitation of the solute obtained either by a $\pi$ pulse of light or by assuming the idealized sudden promotion to the excited state. Moreover, 
we present EOM for the Onsager solvation model and we compare the results with PCM. The developed approach provides qualitatively correct real-time evolutions and is promising as a general tool to investigate the electron dynamics elicited by external electromagnetic fields for molecules in solution.
\end{abstract}

\keywords{Polarizable Continuum Model, electronic excitations, real-time optical response}

\maketitle

\section{Introduction}\label{Sec:Intro}


The study of the electronic dynamics in molecules is a subject of great current experimental interest\cite{Calegari2014} and has been intensively investigated also from a Quantum Chemical perspective.\cite{Kuleff2014} Most of these theoretical works concern the many-electron dynamics in isolated molecular systems neglecting the effects that may be induced by the presence of a dynamically reactive (i.e. polarizable) environment, as a solvent.  In fact, when the molecule is in solution (or in even more complex environment), the electronic dynamics is coupled to the behavior of the solvent, that is in general delayed resulting in a far-from trivial dynamics of the overall solute+solvent system. Within the Polarizable Continuum Model (PCM)\cite{Tomasi2005} for  solvation the general theoretical framework to investigate such dynamics was set long ago,\cite{Cammi1996,Cammi1998,Cammi2013} but only  in a recent paper,\cite{Corni2014} we have shown a convenient way to take into account such coupled dynamics {\it in 
real time}. This approach was based on the definition of {\it equations of motion} (EOM)\cite{Basilevsky1998} for the time-dependent apparent charges that in PCM generate the time-dependent reaction field.\cite{Ingrosso2003,caricato2005time,caricato2006formation} Alternative approaches to study real-time solute electron dynamics within the PCM have also been devised and applied.\cite{liang2012,nguyen2012,Pipolo2014,Ding2015} 

In our previous work, the evolution of the electrons of the molecule was described by a real-time time-dependent density functional theory (RT-TDDFT) approach. While the set-up exploited there  was suitable to show how apparent charge EOM derived for the Debye dielectric constant\cite{bottcher1973theory} could be seamlessly added to the bare RT-TDDFT propagation, it is not ideal to study qualitatively the effect of the solvent on the solute dynamics involving excited states. In fact, it is well-documented that the present versions of RT-TDDFT suffer from various artifacts when excited and/or non-stationary states are in focus: for example, the response to $\pi$-pulses (i.e., light pulses that take all the molecular population from ground state to a given excited state) is unphysical, \cite{Raghunathan2011,Raghunathan2012,Raghunathan2012a} and Rabi oscillations are not properly described.\cite{Habenicht2014} All these artifacts are thought to be consequences of the adiabatic approximation, i.e., the 
assumption that the exchange-correlation (xc) potential at a given time depends only on the density at that time rather than to all the previous instants.\cite{Elliott2012,Provorse2016} 
Although in perspective the use of RT-TDDFT promises to achieve the prominent role that DFT and linear response TDDFT have for investigating ground state {\color{red} properties} and excitation energies and spectra, at present studying electronic dynamics under generic electromagnetic perturbations with RT-TDDFT is complicated by such artifacts. An alternative to RT-TDDFT is the use of time-dependent methods based on the polyelectronic time-dependent Schr\"odinger equation. In practice, for molecules in vacuo this boils down to precalculate a sufficiently large number of excited states energies and dipole elements,  and then to numerically evolve the expansion coefficients of the polyelectronic wavefunction on the eigenstate basis under the time-dependent electromagnetic perturbation. Various groups used this approach in the past for investigating electromagnetic field effects on molecules in vacuo. Krause, Klamroth and Saalfrank\cite{krause2005time} extended the Configuration Interaction singles (doubles), 
CIS(D), to the time-dependent domain (TD-CIS(D))to directly study the laser-pulse-driven many-electron dynamics, focusing in particular on $\pi$ pulses and consequent dipole-switching. Klamroth\cite{Klamroth2006} exploited TD-CIS coupled to the optimal control theory to find the laser pulse shaped that maximizes the population transfer to a selected excited state, and Krause and Klamroth\cite{Krause2008} investigated the role of the molecular size in the selectivity of such pulses. Schlegel, Smith  and Li\cite{Schlegel2007} compared the behavior of TD Hartree Fock (TD-HF), TD-CIS and TD-CIS(D) to describe the dynamics of molecules in intense electromagnetic fields. They found similar behaviors between TD-HF and TD-CIS for low applied electromagnetic fields, while they found differences when the perturbative regime (i.e., excited state populations much smaller than 1) is left. The comparison has been recently extended to EOM-CCSD as well.\cite{Sonk2011,Huber2011} Other wave-function approaches have been based 
on time-dependent self-consistent field methods, \cite{Kato2004,li2005time,Nest2005,Remacle2007} that however are less straightforward in terms of qualitative behaviors due to the non-linearity associated with the self-consistency.\cite{Schlegel2007} 

In the present work, we couple an approach based on a time-dependent polyelectronic wave function with a PCM description of the time-dependent solvent response in the EOM formalism. Thanks to this development, the evolution of the molecular wavefunction in solution under an external perturbation can be studied  free from qualitative artifacts. In this work, we have chosen to rely on the simplest of the wave-function based approaches for excited states, i.e., CIS. Despite its well-known quantitative limitations (in particular its tendency to overestimate excitation energies)\cite{Dreuw2005} it is the simplest method that encompasses the formal features common to more complex approaches, as well as a correct qualitative behavior for solvent-affected excitations such as charge-transfer ones. TD-CIS has been used before to investigate the time-dependent behavior of molecules subjected to electromagnetic radiation.\cite{krause2005time,
Schlegel2007}

The paper is organized as follows: in section \textit{Theory}, the basic concepts of the EOM PCM approach, on one side, and TD-CIS for molecules in vacuo are recalled and the way such approaches can be coupled is described in details. In section \textit{Results and Discussion}, we shall show results for the solvent relaxation following a transition to a given excited state both within the Onsager solvation model and the PCM approach. In particular, we shall create this excited state in two different ways. First we shall simply assume an instantaneous excitation from the ground to the excited state, in the usual fashion of non-equilibrium approaches. This is the standard way the excitations are modeled in the frequency domain calculations of optical properties of molecules in solution.\cite{Cammi2005} Then we shall create the same state by applying a $\pi$ pulse to the molecule, of short but finite length. This is a more physical realization of the excitation, that is possible only when a real-time 
description of the system is at hand. Finally we shall draw the conclusions.

\section{Theory}\label{sec:theory}
We consider a molecular solute involved in a general time-dependent process while is interacting with an external medium (the solvent). The latter is treated as continuum described by a complex dielectric permittivity $\epsilon(\omega)$.  Within PCM,the effective time-dependent electronic Hamiltonian for the solute may be written as:\cite{Cammi1996,Cammi2013}
\begin{equation}
 \hat{H}=\hat{H}^0+{{\v q}}(\Psi;t)\cdot\hat{\v V}+\hat{V}'(t)
\label{eq:H(t)}
\end{equation}
Here, $\hat{H}^0$ is the Hamiltonian of the isolated molecule, ${ {\v q}}(\Psi;t)\cdot\hat{\v V}$ is the potential energy term representing the solute-solvent electrostatic interaction, and $\hat{V}'(t)$ is a general time-dependent external perturbation (we assume that $\hat{V}'(t)$  is applied adiabatically so that it vanishes at $t=-\infty$). 

In Eq. (\ref{eq:H(t)}), ${\v q}(\Psi;t)$ represents the time-dependent solvent polarization charges induced by the solute and placed on the boundary of the cavity hosting the solute itself within the dielectric medium representing the solvent, the dot represents a vector inner product, and $\hat{\v{V}}$ is a vector operator representing the electrostatic potential of the solute at the representative points on the cavity boundary.  
For  general time-dependent processes the time-dependent polarization charges $ {{\v q}}(\Psi;t)$ at a given time $t$ are a function of the previous history of the system:
\begin{equation}
{ {\v q}}({\Psi};t)=\int_{-\infty}^{+\infty}{\v Q}^\t{PCM}(t-t') \langle \Psi(t')|\hat{\v V}|\Psi(t')\rangle dt' 
\label{Q(t)}
\end{equation}  
where ${\v Q}^\t{PCM}(t-t')$ is the solvent response matrix, non-local in time and depending on the whole spectrum of the frequency-dependent dielectric permittivity of the medium.

The state vector $|\Psi(t)\rangle $ of the system satisfies the time-dependent non-linear Schr\"{o}dinger equation
\begin{equation}
 i\frac{\partial}{\partial t}|\Psi(t)\rangle =\hat{H}|\Psi(t)\rangle 
\label{TDNLSE}
\end{equation}
and approximated solutions can be expressed in terms of a many-electron basis set constituted by a suitable reference state $|\Phi_0\rangle $ and the corresponding excited states $|\Phi_I\rangle $ 
\begin{equation}
 |\Psi\rangle = \sum_{I} C_I(t)|\Phi_I\rangle 
 \label{eq:base}
\end{equation}
where $C_I(t)$ are the time-dependent expansion coefficients, and the excited states basis functions ($I> 0$) are generated from the application of an excitation operator $\hat{R}_I$ on the reference state $|\Phi_0\rangle $
\begin{equation}
 \hat{R}_I= \sum_{ijab}\left( r_i^a \hat{a}^\dag\hat{i}+r_{ij}^{ab} \hat{a}^\dag\hat{i}\hat{b}^\dag\hat{j}+...\right)
\end{equation}
Here, $r_{i...}^{a...}$ are the excitation amplitudes of the creation {\color{red}{($\hat{a}^\dag$,$\hat{b}^\dag$,...)}} and annihilation {\color{red}{($\hat{i}$,$\hat{j}$,...)}} operators that generate excitations by promoting electrons from the occupied orbitals $\left\lbrace ij..\right\rbrace$  to vacant orbitals 
$\left\lbrace ab..\right\rbrace$.

Introducing the expansion basis set, Eq. (\ref{eq:base}), the time-dependent Schr\"{o}dinger equation, Eq. (\ref{TDNLSE}), takes the form of a differential equation for the time dependent expansion coefficients 
 \begin{equation}
  i \frac{d {\v C}}{dt}={\v H} {\v C}
  \label{eq:schro}
 \end{equation}
where the Hamiltonian matrix, driving the time evolution, has elements
\begin{equation}
 H_{IJ}(t)=\langle \Phi_I|\hat{H}^0+ {  {\v q}}({\v C};t)\cdot\hat{\v V} +\hat{V}'(t)|\Phi_J\rangle   
\end{equation}
and the time-dependent apparent charges ${  {\v q}}({\v C};t)$ are given by
\begin{equation}
{ {\v q}}({\v C};t)=\int_{-\infty}^{+\infty}{\v Q}^\t{PCM}(t-t') { {\v V}}({\v C};t')dt'
\end{equation} 
with the expectation values of the electrostatic potential expressed as
\begin{equation}
  {{ {\v V}}}({\v C};t) = \sum_{I,J}^NC_I^*(t)C_J(t)\langle \Phi_I| \hat{\v V} |\Phi_J\rangle 
\end{equation}

The accuracy of the solution depends on the nature of the many-electron basis set expansion (Eq. (\ref{eq:base})). For reasons of computational simplicity and feasibility,  we use a reference state given by the Hartee-Fock single determinant $|\Phi_0\rangle =|\t{HF}\rangle $ of the molecular solute, under a regime of equilibrium solvation,  and the excited states $|\Phi_I\rangle $ (with $I> 0$)  in the form of a configuration interaction expansion limited to single excitations (CIS).

The occupied and vacant molecular orbitals associated to the Hartree-Fock wavefunction  $|\t{HF}\rangle $ are obtained from  the solution of the Hartree-Fock equations involving the following Fock matrix (in the molecular orbitals basis):
\begin{equation}
F_{pq}^\t{PCM}=F_{pq}^{0}+{ {\v q}}({|\t{HF}\rangle })\cdot{\v V}_{pq}
\label{eq:fock}
\end{equation}
Here, $F_{pq}^{0}$ are the matrix elements of the Fock operator for the isolated system, and ${ {\v q}}({|HF\rangle})$  collects the apparent  charges induced by the Hartree-Fock charge distribution:
\begin{equation}
{ {\v q}}({|\t{HF}\rangle })= {\v Q}_0 \langle \t{HF}|\hat{\v V}|\t{HF}\rangle ={\v Q}_0 { {\v V}}({|\t{HF}\rangle })
\end{equation}
where the response matrix $ {\v Q}_0 $ is in the regime of equilibrium solvation, and depends on the static dielectric permittivity $\epsilon_0 $ of the medium.
${\v V}_{pq}$ are the elements of the representation matrix of the vectorial operator ${\v V}$ in the MO basis.

The amplitudes  and energies of the CIS excited states $|\Phi_I\rangle $  are obtained by solving, in the {\color{red} space spanned by the Hartree-Fock determinant and by the}  single excited determinants $|\Psi_i^a\rangle =\hat{a}^\dag\hat{i}|\t{HF}\rangle  $, the time-independent Schr\"{o}dinger equation for  the molecular solute in the presence of the fixed Hartee-Fock polarization charges:
\begin{equation}
\left [ \hat{H}^0+{  {\v q}}(|\t{HF}\rangle ;t)\cdot\hat{\v V} \right ]|\Phi_I\rangle =E_I|\Phi_I\rangle 
\label{eq:ei}
\end{equation}
The corresponding Hamiltonian matrix to be diagonalized has elements:
\begin{eqnarray}
 \langle \t{HF}|\hat{H}^0+{  {\v q}}(|\t{HF}\rangle)\cdot\hat{\v V} |\Psi_i^a\rangle =0 \\
 \langle \Psi_i^a|\hat{H}^0+{  {\v q}}(|\t{HF}\rangle)\cdot\hat{\v V} |\Psi_j^b\rangle =H_{ij}^{ab}
\end{eqnarray}

 {\color{red} The reference states $|\Phi_I\rangle $ ($I>0$) defined by Eq. (12)  correspond to the the excited states of the solute in the frozen-solvent approximation, i.e., by assuming that upon the solute excitation, the solvent reaction field remains the same as in the ground state. This is the same reference state choice made in our previous work dealing with state-specific vs linear-response treatments of excitations in solution.\cite{Cammi2005}$^,$
 \footnote{{\color{red} In principle, other reference states could be used, such as the eigenstates for the gas-phase molecule. However, none of such gas-phase states is stationary for a molecule in solution, making the analysis of the perturbation-induced dynamics unreliable.}}
} Using the Hartree-Fock state {\color{red} ($I=0$)} and the CIS excited states {\color{red} ($I>0$)} as an expansion basis set, the elements of the Hamiltonian matrix that drives the time evolution of the state vector (Eq. (\ref{eq:H(t)})) take the form:
\begin{eqnarray}
H_{IJ}(t)&=&E_I\delta_{IJ}+ \langle \Phi_I|\Delta{  {\v q}}({\v C};t)\cdot\hat{\v V} +\hat{V}'(t)|\Phi_J\rangle
\label{eq:H_IJ}
\end{eqnarray}
where the time-dependent apparent charges $\Delta{  {\v q}}({\v C};t)$ are given by
\begin{equation}
{\Delta {\v q}}({\v C};t)= {\v q}({\v C};t)-  {\v q}(|\t{HF}\rangle)=\int_{-\infty}^{+\infty}{\v Q}^{\t{PCM}}(t-t') {\Delta {\v V}}({\v C};t')dt'
\label{eq:int}
\end{equation} 
with 
\begin{equation}
{\Delta {\v V}}({\v C};t')={ {\v V}}({\v C};t')-{ {\v V}}({|\t{HF}\rangle })
\label{eq:deltaq}
\end{equation}
{\color{red} In general, the Hamiltonian in Eq.(\ref{eq:H_IJ}) is diagonal only when $\Delta{  {\v q}}({\v C};t)=0$ (i.e., the reaction field is that in equilibrium with the ground state) and there is no external perturbation. Therefore, by starting a dynamics from one of the frozen-solvent excited states $|\Phi_I\rangle $, such state will mix with the others, including the ground state.  } 

Here, we note that the set of Eqs. (\ref{eq:schro}-\ref{eq:deltaq}) constitutes the basic formulation of the time-dependent CIS method for a molecule in solution. However, in this basic formulation the time evolution of the polarization charges of the solvent is expressed in terms of an integral representation Eq. (\ref{eq:int}) which is not convenient within a real-time approach to the solution of the time-dependent Schr\"{o}dinger equation (\ref{eq:schro}).

\subsection{Equation-of-motion for the solvent polarization charges }

We now consider the case of a solvent whose frequency dependent dielectric permittivity $\epsilon(\omega)$ may be  approximated by the Debye relation
\begin{equation}\label{eq:Deb}
 \epsilon(\omega)=\epsilon_{d}+\frac{\epsilon_{0}-\epsilon_{d}}{1-i\omega\tau_D}
\end{equation}
where $\omega$ is the frequency, $\epsilon_0$ and $\epsilon_{d}$ are the static and optical frequency dielectric permittivities, respectively, and $\tau_D$ is the Debye relaxation time of the solvent. The time-dependent polarization charges in Eq. \ref{eq:deltaq} are given by:\cite{Corni2014}
\begin{eqnarray}\label{eq:dqpcm_int}
 \Delta {{ {\v q}}}({\v C};t)&=&\v{Q}_d \cdot \Delta {{ {\v V}}}({\v C};t)
  +\int_{-\infty}^{\infty} dt'\v{Q}_{\t{D}}^{\t{PCM}}(t-t') \Delta{ {\v V}}(\v{C};t')
\end{eqnarray}
The forms of the Debye's kernel matrix $\v{Q}_{\t{D}}^{\t{PCM}}(t-t')$ and the optical response matrix $\v{Q}_d$ are different for the different formulations of the PCM. In this work we use the diagonal formulation of the Integral Equation Formalism, IEF(d) in ref. \citenum{Corni2014}.
A direct time-differentiation of Eq. (\ref{eq:dqpcm_int}) gives the following equation of motion for the apparent charges (Debye EOM TD-PCM):
\begin{eqnarray}\label{eq:dqpcm}
 \frac{d }{dt}\Delta {{ {\v q}}}(\v{C};t)&=&\v{Q}_d \frac{d }{dt}\Delta{{ {\v V}}}({\v C};t)+\v{\tilde{Q}} { {\v V}}({\v C};t)- \v{R} \Delta{ {\v q}}(\v{C};t)
\end{eqnarray}
The matrices $\v{\tilde{Q}}$ and $\v{R}$ have been defined in ref.\citenum{Corni2014}.
Eq. (\ref{eq:dqpcm}) for the polarization charges and the time-dependent Schr\"{o}dinger equation, Eq. (\ref{eq:schro}), constitute a system of two equations of motion describing the many-electron dynamics of the molecular solute under the influence of the solvent polarization and viceversa. This system of differential equations can be integrated by using a simple Euler method, according to which the dynamic variables (expansion coefficients and polarization charges) a time $t+\Delta t$  are given by
\begin{eqnarray}
 i{\v C}(t+\Delta t) &=& {\v H}(\Delta {{ {\v q}}};t){\v C}(t)\Delta t + i{\v C}(t) \label{eq:CDeltat} \\
 \Delta {{ {\v q}}}({\v C};t+\Delta t)&=&  \v{Q}_d \left[ \Delta {{ {\v V}}}({\v C};t+\Delta t)-\Delta {{ {\v V}}}({\v C};t)\right]+ \Delta t ~\v{\tilde{Q}}~\Delta {{ {\v V}}}({\v C};t)+ \nonumber \\
   & & - \Delta t ~\v{R}~\Delta {{ {\v q}}}({\v C};t) + \Delta {{ {\v q}}}({\v C};t)
\end{eqnarray}

In Eq. (\ref{eq:CDeltat}) the dependence of the Hamiltonian matrix elements, defined in Eq. (\ref{eq:H_IJ}), on the time-dependent polarization charges is explicitly shown.

We conclude this section by noting that similar EOMs can be obtained for the Onsager solvation model\cite{onsager1936electric,van1982dynamical,van1985time} (Debye EOM TD-Onsager) in terms of a time dependent reaction field $\Delta\vec{F}$ and a dipole moment variation ${\Delta\vec{\mu}}$:
\begin{eqnarray}\label{eq:df_ons}
 \frac{d}{dt}\Delta{\vec{F}}(\v{C};t)&=&g_d  \frac{d }{dt} \Delta{\vec{\mu}}({\v C};t) +\frac{g_0}{\tau_{\t{Ons}}} \Delta{\vec{\mu}}({\v C};t)- \frac{1}{\tau_{\t{Ons}}} \Delta{\vec{F}}(\v{C};t)
\end{eqnarray}
where $g_0$ and $g_d$ are the static and dynamic reaction field factors respectively.\cite{bottcher1973theory}
$\Delta\vec{F}$ and ${\Delta\vec{\mu}}$ are defined as follows:
\begin{eqnarray}
\nonumber \Delta\vec{F}({\v C};t)&=&\vec{F}({\v C};t)-\vec{F}(|\t{HF}\rangle)\\
&=&g_0 ~\Delta\vec{\mu}(t) + \int_{-\infty}^t dt' \left [\frac{e^{-(t-t')/\tau_{\t{Ons}}}}{\tau_{\t{Ons}}} (g_0-g_d) \right] \Delta\vec{\mu}({\v C};t')\\
\Delta\vec{\mu}({\v C};t')&=&\vec{\mu}({\v C};t')-\vec{\mu}({|\t{HF}\rangle })
\end{eqnarray}
with $\tau_{\t{Ons}}={\tau_\t{D}(2\epsilon_d+1)}/{(2\epsilon_0+1)}$.

Extension to higher-order multipoles and ellipsoidal cavities is straightforward, the equations are reported in Appendices \ref{App-Mult} and \ref{App-Ell}, respectively. 

\subsection{Non-equilibrium time-dependent free energy}
It is well-known that in a time-dependent picture the equilibrium definition of solvation free energy should be extended.\cite{Bonaccorsi1983,Aguilar1993,Cammi1995} In particular, we need here a definition of non-equilibrium free energy, $\mathcal{G}^{neq}$, suitable for a time-dependent solvation. Such free energy has been defined by Caricato et al.\cite{caricato2006formation} By reformulating $\mathcal{G}^{neq}$ with the quantities and nomenclature of the present work we obtain:
\begin{eqnarray}
\mathcal{G}^{neq}(t)= {E}(t)+\frac{1}{2}  {\v q}_d(t) \cdot  {\v V}(t)+ ( {\v q}(t)- {\v q}_d(t))\cdot  {\v V}(t)-\frac{1}{2}( {\v q}(t)- {\v q}_d(t))\cdot \tilde{\v V}(t)
\label{eq:gneq}
\end{eqnarray}
where $ {E}(t)= \sum_{I,J}C_I^*(t)C_J(t)<\Phi_I| \hat{H}^0+\hat{V}'(t) |\Phi_J>$ is the expectation value of the gas-phase Hamiltonian including the perturbation, $ {\v q}_d(t)$ are the charges obtained considering the instantaneous  response of the solvent to the solute charge density at the time $t$ and $\tilde{\v V}(t)$, already defined in ref.\citenum{caricato2006formation}, is the molecular potential that would produce the charges $ {\v q}(t)$ via the solvent static response. The dependence of charges and potentials on the coefficients of the CIS expansion is not shown explicitly in this section for the sake of clarity. The last term in eq.(\ref{eq:gneq}) accounts for the free-energy cost of creating the inertial polarization represented by $ {\v q}(t)- {\v q}_d(t)$. 
$\mathcal{G}^{neq}(t)$ encompasses the non-equilibrium free energy expression after a sudden transition as a special case, and converges to the standard equilibrium free energy in the long time limit.\cite{caricato2006formation}

The absolute value of $\mathcal{G}^{neq}(t)$ is not, however, particularly useful in the analysis of the time evolution of the system, since it is always defined up to an arbitrary constant. Rather, the {\it de-excitation energy} $\Delta G^{neq}_{dex}(t)$ at the time $t$, i.e., the non-equilibrium free-energy variation in suddenly passing from the excited state to the ground state, is more useful.  ``Suddenly'' in this context means that the time-dependent solvent degrees of freedom are assumed to be frozen during the transition, and only the part represented by $\epsilon_d$ can instantaneously follow the solute electronic transition 
\begin{eqnarray}
\nonumber 
-\Delta G^{neq}_{dex}(t)&= & \mathcal{G}^{neq}(t)-\mathcal{G}^{neq}_{0}(t)\approx \\
\nonumber 
&=& {E}(t)-\langle \t{HF}|\hat{H}^0|\t{HF}\rangle
+\frac{1}{2} ( {\v q}_d(t) \cdot  {\v V}(t)- {\v q}_{d}(|\t{HF}\rangle) \cdot  {\v V}(|\t{HF}\rangle))+ \\&&( {\v q}(t)- {\v q}_d(t))\cdot( {\v V}(t)- {\v V}(|\t{HF}\rangle))
\label{eq:dgneq}
\end{eqnarray}
Using $|\t{HF}\rangle $ as the ground state wavefunction at each time $t$ is actually an approximation: since the solvent polarization is changed during the evolution in the excited state, the ground state wavefunction after the de-excitation is different from $|\t{HF}\rangle $. It is possible to recalculate such wavefunction, but it would be too expensive to be obtained at each step of the evolution. The numerical results presented later are therefore obtained within this approximation.

Eq.(\ref{eq:dgneq}) can be simplified when the Onsager model is used:
\begin{eqnarray}
\nonumber 
-\Delta G^{neq}_{dex,Ons}(t)= {E}(t)-\langle \t{HF}|H^0|\t{HF}\rangle
-\frac{1}{2} g_d (|\vec{\mu}|^2(t)-|\vec{\mu}(|\t{HF}\rangle)|^2) + \\ - 
(\vec{F}(t)-g_d \vec{\mu}(t)) \cdot ( \vec{\mu}(t)-\vec{\mu}(|\t{HF}\rangle)))
\label{eq:dgneq_ons}
\end{eqnarray}

We close this section by remarking that when the molecule undergoes a sudden excitation to an excited state at $t=t_0$, then $-\Delta G^{neq}_{dex}(t_0)$ is the non-equilibrium excitation energy in solution, and the values for $t>t_0$ are estimates of the time-dependent fluorescence energy  ( from which time-dependent Stokes shifts can be recovered).\cite{Ingrosso2003}

\section{Results and Discussion}
In this section we present the numerical results of the EOM-PCM/TD-CIS method presented above. The aim is to check the physical consistency of the method, by comparison with the corresponding Onsager model, and to show some of the potentialities of the method. For these reasons, and for the sake of simplicity, we have considered a model system, LiCN, which was previously used for TD-CIS calculation in vacuo,\cite{krause2005time} in acetonitrile.

\subsection{Computational Details}
The geometry of the LiCN molecule and the level of theory used for the calculations are the same as in ref.\citenum{krause2005time}: the length of the LiC and CN bonds are respectively set to $1.949$ \AA{} and $1.147$ \AA{}, the 6-31G(d) basis set\cite{hariharan1973influence} is used.
The molecular-shaped cavity is generated using the GePol algorithm \cite{pascual1994gepol} using three spheres centered at atomic position with radii equal to the atomic van der Waals radii reported in ref.\citenum{mantina2009consistent} ($r_{Li}=1.81$ \AA{}, $r_{C}=1.70$ \AA{}, $r_{N}=1.55$ \AA{})
scaled by a factor of $1.2$.
The parameters of the Debye's dielectric function for acetonitrile are taken from ref.\citenum{caricato2005time} ($\epsilon_0 = 35.84$, $\epsilon_d=1.806$, $\tau_D = 3.37$ps). {\color{red} For the investigation of the sudden excitation energies in solution and of the excited state populations following $\pi$-pulses, both this value of $\epsilon_d$ and the approximation $\epsilon_d=1.0$ will be employed. Only} $\epsilon_d=1.0$ will be used {\color{red} in all the other} simulations. 
The time step for the numerical propagation is $dt = 4.838~$as ({\color{red} $0.2~$}au). 
The calculation of the \t{HF} and CIS states {\color{red} in vacuum and} in the presence of the PCM solvent, their energies (at frozen solvent reaction field), the expectation values and transition integrals of the dipole moment and the electrostatic potential operators, are performed with a locally modified version of the Gaussian G09 (A02) package. \cite{G09trunc} In {\color{red} the } numerical applications  {\color{red} of the EOM-PCM/TD-CIS method}  we have limited the many-electron basis set to the  {\color{red} Hartree-Fock}  ground state and to  {\color{red} the lowest} 5 CIS excited states {\color{red} determined in the presence of the PCM frozen reaction field.}  {\color{red} Energies and dipole expectation values for this many-electron wavefunction basis set are reported in Table \ref{tab:states}. A complete analysis of the lowest 5 CIS states in terms of Slater determinants and symmetries is reported in the Supplementary Material section.} This choice {of the many-electron basis set} allows reducing the mixing of the electronic states during the time-dependent evolution, which can lead to complex electronic processes (that will be the subject of future works), and to rather focus on the solvent dielectric relaxation. {\color{red} Generally speaking, this number of state is 
small and, taking also into account the characteristics of the pulse, not enough to achieve a converged result with respect to the reference basis set. In applications targeting specific experiments, tests on the convergence with the number of states are needed.}
The coupled propagation of the wavefunction of the molecule and of the reaction field is performed using WaveT, an home-made code.

\begin{table}
{\color{red}{
\caption{Energies ($E_{I}$ in eV) and dipole expectation values along the molecular axis ($\mu_{\t{II},z}$ in Debye) for the ground (S$_0$) and excited (S$_I$, $I>0$) calculations in vacuum (Vacuum) and with PCM at frozen solvent reaction field (PCM frozen). Note that the nature of the lowest excited states may change in passing from the vacuum to the PCM solvent (see Supplementary Material section). }
\label{tab:states}
\begin{tabular}{|c|cc|cc|}
\hline
\multirow{2}{*}{State} &\multicolumn{2}{c|}{Vacuum}&\multicolumn{2}{c|}{PCM frozen}\\
\cline{2-5}
 & $E_I$ &$\mu_{II,z}$& $E_I$ &$\mu_{II,z}$\\
\hline
S$_0$ & 0.0     &  -9.4250 & 0.0     &  -10.6814 \\
S$_1$ & 6.50260 &  -4.0254 & 7.27854 &   -7.6306 \\
S$_2$ & 6.57988 &   7.1044 & 7.55445 &   -7.2069 \\
S$_3$ & 6.57988 &   7.1044 & 7.55469 &   -7.2069 \\
S$_4$ & 6.74522 &  -2.9932 & 8.31361 &   -7.2292 \\
S$_5$ & 6.74522 &  -2.9932 & 8.31361 &   -7.2292 \\
\hline
\end{tabular}
}}
\end{table}

\subsection{Relaxation of the solvent after a sudden excitation: Onsager vs PCM}
\begin{figure}
\centering  \includegraphics[width=0.6\textwidth]{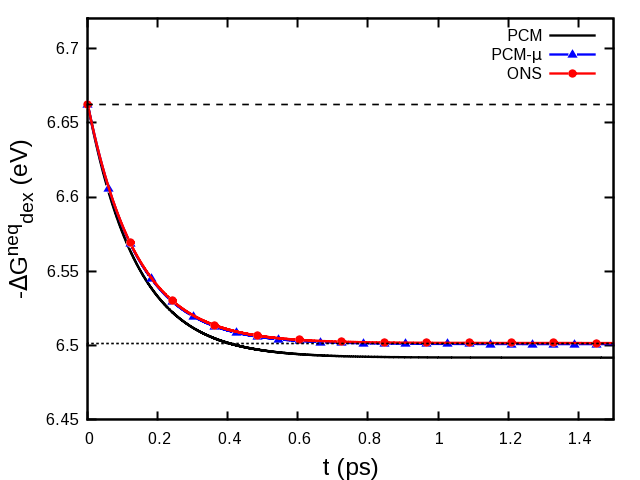} 
\caption{$\Delta G^{neq}_{dex}(t)$ for LiCN in a spherical cavity after a sudden excitation from the ground state to the state S$_4$. The cavity has a radius of $10${\AA}. ``PCM'' refers to the results from a simulation using PCM; ``PCM$-\mu$'' refers to a PCM simulations in which the {\color{red} molecular electrostatic} potential of the solute is calculated from the molecular dipole only, rather than from the entire density; ``ONS'' refers to a TD simulation based on the Onsager model.
Horizontal thin dashed and dotted lines show the values of $\Delta G^{neq}_{dex,Ons}(t)$ just after the sudden jump (dashed) and after the Debye's relaxation (dotted), calculated as limit cases of Eq. (\ref{eq:dgneq_ons}). {\color{red} Note that the ``PCM$-\mu$'' and the ``ONS'' curves are on-top of each-other.}}
\label{fig:ons_vs_pcm} 
\end{figure}

As a first test of our approach, we have considered the solvent relaxation following a sudden excitation to the first dipole-allowed ground state{\color{red}{. Such a state, S$_4$, is degenerate with S$_5$, as shown in Table \ref{tab:states} and the expectation value of its dipole moment is parallel to the one of the ground state but with a smaller modulus. 

In Fig.(\ref{fig:ons_vs_pcm}) we compare the non-equilibrium free-energy de-excitation traces $\Delta G^{neq}_{dex}(t)$ obtained with Debye EOM TD-PCM  with that coming from Debye EOM TD-Onsager.}} 
The agreement between the Onsager and the PCM results is very good just after the transition (i.e., at time $t_0$=0 ps), while at the end of the simulation a minor discrepancy of about 10 meV (over a solvatochromic shift relaxation of $\approx$170 meV) is present. This can be attributed to the role of solute multipole moments higher than the dipole. In fact, when such multipole moments are artificially switched off (``PCM$-\mu$'' in Fig.(\ref{fig:ons_vs_pcm})), the PCM and Onsager results perfectly agree.
\subsection{Solvent relaxation after an excitation: sudden vs $\pi-$pulse preparation}

In the previous section, the starting point of the Debye EOM TD-PCM/TD-CIS simulation was created by a theoretical idealization, i.e., a sudden electronic excitation of the solute during which the slow degrees of freedom of the solvent are frozen. Here we shall focus again on the same excited state and the solvent relaxation taking place after an excitation to such state, but this time the state is produced in a physically realistic way, i.e., by a light-pulse able to take the ground state to the target excited state. Such light-pulse, called $\pi$ pulse in analogy with NMR spectroscopy, is characterized by {\color{red} specific time dependent electric fields. For molecules in gas-phase, modelled as a two-level system and in the rotating wave approximation, one of the possible time-dependence} reads:\cite{krause2005time}

\begin{eqnarray}
\vec{F}_{\pi}(t)=\begin{cases}
  \vec{\mu}_{0I}\frac{\pi}{\sigma_p \mu_{0I}^2}\cos^2\left( \frac{\pi}{2 \sigma_p} (t-t_p) \right)~\cos\left(\omega_{0I}(t-t_p)\right), & \t{if } |t-t_p|< \sigma_p.\\
  0, & \t{otherwise}.
  \end{cases}
 \label{eq:pi}
\end{eqnarray}  

where $\vec{F}_{\pi}(t)$ is the electric field of the $\pi$ pulse, $\vec{\mu}_{0I}$ is the transition dipole from the ground state to the excited state $I$, $\sigma_p$ is the full-width at half maximum of the pulse, $t_p$ is the central time of the pulse and $\omega_{0I}$ is the excitation frequency for the transition  from the ground state to the excites state $I$. {\color{red} Other functional forms for $\vec{F}_{\pi}(t)$ are possible.\cite{Tremblay2008}} The time dependent perturbation in Eq. (\ref{eq:H(t)}) associated to the $\pi$ pulse is defined in the dipole approximation as $V'(t)=-\vec{\mu}\cdot \vec{F}_{\pi}(t)$.

\begin{figure}
\centering  \includegraphics[width=0.6\textwidth]{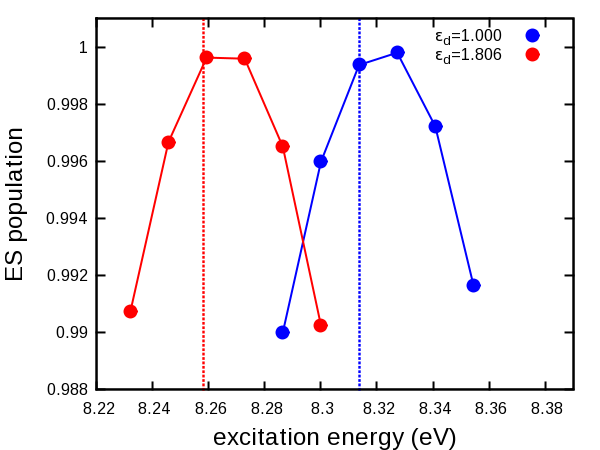} 
\caption{Final populations achieved after the pulse as a function of the excitation {\color{red} energy} $\omega_{0I}$ in Eq.(\ref{eq:pi}). The results for $\epsilon_d=1.000$ and $\epsilon_d=1.806$ are reported. Vertical dashed lines represent the non-equilibrium excitation energies for a sudden excitation to S$_4$. {\color{red} In particular the blue dashed line, corresponding to $\epsilon_d=1.000$, represents the frozen-solvent excitation energy.} }
\label{fig:pop-pulse} 
\end{figure}

\begin{figure}
\centering  \includegraphics[width=0.6\textwidth]{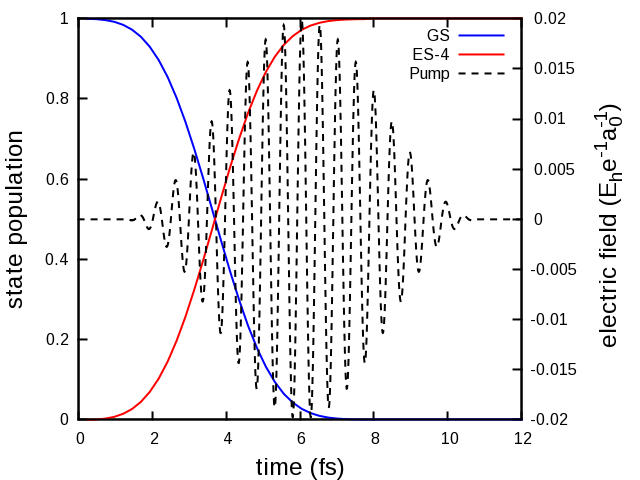} 
\caption{Left axis: Population of the ground (``GS'') and of the target excited state {\color{red} S$_4$ }(``ES-4'') as a function of time during the $\pi$ pulse. Right axis: Electric field component along the $\vec{\mu}_{04}$ direction associated with the $\pi$ pulse.}
\label{fig:pulse} 
\end{figure}

In the present case, we have considered LiCN in acetonitrile solution (with a realistic, molecularly shaped, cavity described in the Computational Details section), we have chosen the lowest bright excited state (S$_4$) as the target state $I$ and the electric field is applied along the $\vec{\mu}_{04}$ direction (see Table \ref{tab:dip}). The chosen excited state is not that with a dipole moment opposite to the ground state one, as previously addressed in gas-phase simulations,\cite{krause2005time} since the latter is shifted to high energies in the unfavorable ground state reaction field (i.e., by the interaction with the solvent). The $\pi$ pulse for the molecule in solution was based on Eq.(\ref{eq:pi}) as well. In particular, we chose $\sigma_p=200~$au$~=4.83~$fs, we used the components of the $\vec{\mu}_{04}$ transition dipole reported in Table \ref{tab:dip} (no local field effects were included) and we optimized the $\omega_{04}$ excitation energy by a trial-and-error procedure.
\begin{table}
{\color{red}{
\caption{Non-zero transition dipole moments ($\vec{\mu}_{IJ}$) among States S$_I$ and S$_J$ calculated using PCM at frozen solvent reaction field. Values in Debye.}
\label{tab:dip}
\begin{tabular}{|c|ccc|}
\hline
States S$_I$ - S$_J$ &$\mu_{IJ,x}$&$\mu_{IJ,y}$&$\mu_{IJ,z}$\\
\hline
S$_0$ - S$_4$  &   0.2916  &    0.1085  &   0.0000 \\
S$_0$ - S$_5$  &   0.1085  &   -0.2916  &   0.0000 \\
S$_4$ - S$_1$  &   0.0653  &   -0.1755  &   0.0000 \\
S$_4$ - S$_2$  &  -0.0049  &    0.0018  &   0.0000 \\
S$_4$ - S$_3$  &   0.0018  &    0.0048  &   0.0000 \\
S$_5$ - S$_1$  &  -0.1755  &   -0.0653  &   0.0000 \\
S$_5$ - S$_2$  &  -0.0018  &   -0.0049  &   0.0000 \\
S$_5$ - S$_3$  &  -0.0048  &    0.0018  &   0.0000 \\
\hline
\end{tabular}
}}
\end{table}
We considered a few tentative values of $\omega_{04}$ close to the expected excitation frequency $-\Delta G^{neq}_{dex}(t_0)$, i.e., the non-equilibrium excitation energy, monitoring the final population achieved after the pulse. The results are shown in Fig.\ref{fig:pop-pulse}.
The excitation frequency providing the higher population transfer is very close to the expected one (not surprisingly since $\sigma_p$ chosen here, $\approx 5$fs is very small compared to the acetonitrile relaxation time, $\tau_D=3.37$ps), and the $\pi$ pulse transfer more than 99\% of population to S$_4$. The electric field associated to such pulse as well as the populations of the ground and the fourth excited states as a function of the time during the pulse are reported in Fig. (\ref{fig:pulse}). 

\begin{figure}
\centering  \includegraphics[width=0.6\textwidth]{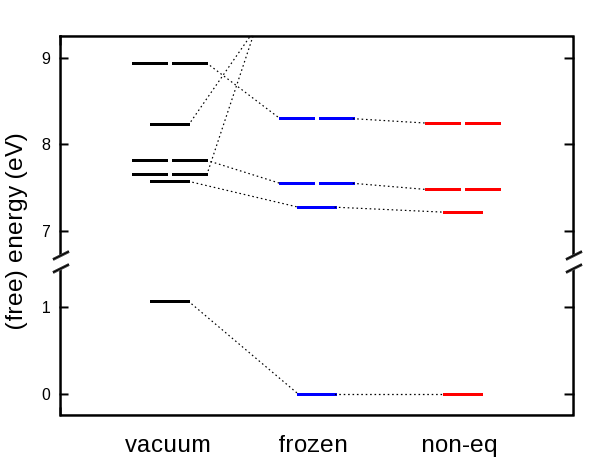} 
\caption{{\color{red} Energies (gas-phase) and Free energies (in solvent) of the ground and the lowest excited states for the molecule in gas-phase (vacuum - left), within the frozen solvent approximation (frozen - center) and with the non-equilibrium approach just after a sudden excitation (non-eq - right). Dashed lines connect states of the same nature. All the values refer to the equilibrium free energy of the ground state in solution, that is therefore set to zero. $E_I$ in eq.(\ref{eq:ei}) correspond to the "frozen" values.}}
\label{fig:corr}
\end{figure}

{\color{red} In Fig.(\ref{fig:pop-pulse}), we report the results for two choices of the $\epsilon_d$ dielectric function: one is that proper for acetonitrile ($\epsilon_d=1.806$), the other ($\epsilon_d=1.000$) corresponds to  a situation where no degrees of freedom of the solvent can instantaneously re-adjust with the solute state. Immediately after a sudden excitation, or a short $\pi$ pulse, this corresponds to a frozen-solvent approximation. As it can be seen in the Figure, including the  instantaneous solvent response (red dotted-line) gives a red-shift with respect to the frozen-solvent excitation. This is due to the stabilization of the excited state  thanks to the instantaneous adaptation of the solvent electronic degree of freedom to the solute excited state wavefunction, that is neglected in the frozen-solvent approximation. To complete the excitation energy analysis, we report in Fig.(\ref{fig:corr}) a correlation plot showing the energies of the lowest states in gas-phase, their non-equilibrium free-energy within the frozen 
solvent approximation and by including also the instantaneous electronic response (but keeping the same solute wavefunctions as in the frozen-solvent approximation). Clearly, the largest contribution of the solvent effect is already at the frozen-solvent level, and stems mainly from the stabilization of the (highly polar) ground state. Coherently, the excited states with a dipole directed opposite to the ground state's are taken to very high energies by the solvent effect.}

\begin{figure}
\centering  \includegraphics[width=0.6\textwidth]{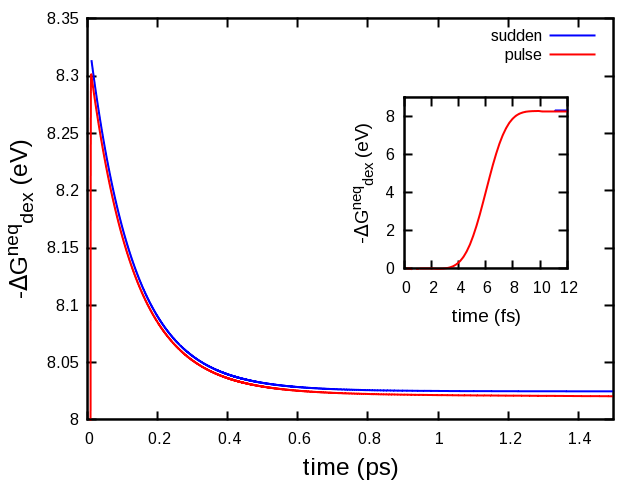} 
\caption{$\Delta G^{neq}_{dex}(t)$ as a function of time for LiCN in a molecular shaped cavity after a sudden excitation to S$_4$ (``sudden'') and after the $\pi-$ pulse of eq.(\ref{eq:pi}) (``pulse''). The inset zooms on the behavior of $\Delta G^{neq}_{dex}(t)$ during the pulse (note the fs time scale).}
\label{fig:sud_vs_pulse} 
\end{figure}

After having determined the parameters of the $\pi$ pulse, we shall then compare the relaxation of the solvent following the theoretical sudden excitation and the $\pi-$pulse excitation. {\color{red} Here $\epsilon_d$ is set to 1.000, which allows using directly the frozen-solvent state as the starting condition for the sudden excitation approach (no such simplification would be required for the $\pi-$pulse excitation).} The deexcitation energy traces are shown in Fig.(\ref{fig:sud_vs_pulse}).
Excluding the few initial fs, the agreement between the curves is very high. Such an agreement is, a posteriori, a justification of the ``sudden'' approximation used ubiquitously in the literature of solvatochromic shifts. {\color{red}{Furthermore for the same simulation we have checked the quality of the approximation made in eq.(\ref{eq:dgneq}): at the end of the simulation the contribution of the HF state to the ground state is higher than 99.99\%.}}

\begin{figure}
\centering  \includegraphics[width=0.6\textwidth]{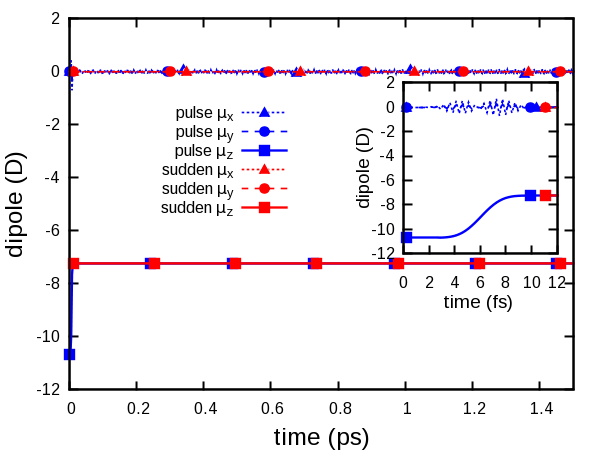} 
\caption{Molecular dipole components as a function of time for LiCN in a molecular shaped cavity after a sudden excitation to S$_4$ (``sudden'') and after the $\pi$ pulse of eq.(\ref{eq:pi})(``pulse''). The inset zooms on the behavior of the dipoles during the pulse (note the fs time scale). {\color{red} Outside the pulse interval, the dipole components $\mu_x$ and $\mu_y$ are zero for both the pulse and the sudden excitations, while their $\mu_z$ superimpose perfectly.}}
\label{fig:dipole} 
\end{figure}

The plot of the time-dependent expectation value of the dipole as a function of time is reported in Fig.(\ref{fig:dipole}).
After a transient change from the ground to the excited state values for the $\pi-$ pulse simulations, the expectation value of the dipole moment becomes constant and equal to that of the state created with the sudden excitation. Only minimal oscillations along $x$ and $y$ directions can be seen in the plot (for the $\pi$-pulse trace), due to a residual ground state component. The constant value of the dipole moment in this plot also indicates that, within the small excited state space used for these tests, the solvent reaction field relaxation is not polarizing the solute wavefunction (i.e., it is not mixing it with other excited state wavefunctions). Also, on the time scale of these simulations, we do not observe decay to the ground state due to solute electronic energy dissipation into the solvent, which is reasonable due to the largely different time-scales that characterize the electronic dynamics of the solute and the nuclear dynamics, embodied in the Debye dielectric constant, of the solvent.

\section{Conclusion}

In this article, we have developed the theory to couple the time-dependent Schr\"{o}dinger equation for a wavefunction-based description of a molecular solute subject to an external electromagnetic radiation with a time-dependent EOM description of the surrounding solvent, within the framework of the PCM continuum solvation model. In particular, we have focused on the simplest possible wave-function description of the excited states (CIS) that, although known for the moderate accuracy of the resulting excitation energies,\cite{Dreuw2005} provides an effective test-bed for the theory. With regard to the solvent, we have used a description based on the Debye dielectric function, that embodies the core of the solvent nuclear relaxation effect. Extension of the EOM TD-PCM method to more complex dielectric functions (e.g., multiple relaxation time Debye, Onodera\cite{Onodera1993,Schwerdtfeger2014}) is possible and will be done in the future.

To illustrate the computational implementation, we have presented results concerning the solvent relaxation after an electronic excitation for the LiCN molecule. In particular, we have compared the EOM TD-PCM results with those obtained for the EOM TD-Onsager solvation model. Moreover, we have shown that the relaxation following an idealized sudden jump from the ground to an excited state is the same as that obtained by taking the molecule to the same excited state by a proper light-pulse. This result backs-up the assumption of the sudden jump, that is at the basis of the current theories of solvatochromic shifts.\cite{Cammi2005}

The theory and the corresponding implementation presented here provide the basis to explore, at least qualitatively, a plethora of time-dependent phenomena, such as absorption by arbitrarily shaped pulses, time-dependent Stokes shifts,\cite{Ingrosso2003} optimal control\cite{Klamroth2006} for molecules in solution, investigation of excited energy dissipation by the solvent and excited state electron transfer dynamics controlled by the solvent, free from the mystifying qualitative artifacts that may plague real-time TDDFT results in the adiabatic approximation.

{\color{red} \section*{Supplementary Material}
See supplementary material for the description of the lowest 5 excited states in terms of single excitations.}

\section*{Acknowledgments}
S.C. acknowledges funding under the ERC Consolidator Grant 681285 TAME-Plasmons.

\appendix
\section{EOM for the Multipole Reaction Field in a Spherical Cavity}
\label{App-Mult}
In this section we present a derivation of the equations of motion (EOM) for the reaction potential in spherical cavities.
We start from the expression for the (multipole) reaction potential given by B\"ottcher \cite{bottcher1973theory} in spherical coordinates ($r,\theta,\psi$) 
and frequency domain:
\begin{align}\label{eq:mult}
V_R(r,\theta,\psi,\omega)&=- \sum_{l=0}^{\infty} r^l \sum_{m=-l}^{l} V_{lm}(\omega)Y_{lm}(\theta,\psi) \\
V_{lm}(\omega)&=K_lf_l(\omega)M_{lm}(\omega)
\end{align}
with
\begin{align}\label{eq:fl-mult}
f_l(\omega)&=\frac{\epsilon(\omega)-1}{\epsilon(\omega)+k_l}
\end{align}
and
\begin{align}\label{eq:Kk-mult}
K_l&=\frac{1}{a^{2l+1}}         & k_l&=\frac{l}{l+1}
\end{align}
here $V_{lm}$ are the frequency-dependent contributions to the reaction potential given by each multipole component $M_{lm}$, $Y_{lm}(\theta,\psi)$ are the associated spherical harmonics, and $a$ is the radius of the spherical cavity.
In the time domain we may write:
\begin{align}\label{eq:mult-t}
V_R(r,\theta,\psi,t)&=- \sum_{l=0}^{\infty} r^l \sum_{m=-l}^{l} V_{lm}(t)Y_{lm}(\theta,\psi) \\
V_{lm}(t)&=K_l\int_{-\infty}^{\infty}f_l(t-t')M_{lm}(t') dt'\label{eq:mult1}\\
f_l(t-t')&=\int_{-\infty}^{\infty}\frac{d \omega}{2\pi} e^{-i\omega (t-t')} f_l(\omega)\label{eq:mult2}
\end{align}
Following the route given in ref. \citenum{Corni2014} we provide the following expression for the time-dependent multipolar contributions to the reaction potential within the Debye's model for the dielectric function (Eq. (\ref{eq:Deb})):
\begin{align}\label{eq:multxi-t}
V_{lm}(t)&=K_l f_{l,d} M_{lm}(t)  + K_l \int_{-\infty}^{\infty} dt' \left [\frac{e^{-(t-t')/\tau_l}}{\tau_l}\Theta(t-t') (f_{l,0}-f_{l,d}) \right] M_{lm}(t') 
\end{align}
furthermore we obtain the following equations of motion for $V_{lm}(t)$:
\begin{align}\label{eq:EOMmult}
 \frac{d}{dt}~V_{lm}(t)=K_lf_{l,d}\frac{d}{dt} ~M_{lm}(t)+\frac{K_lf_{l,0}}{\tau_l}~M_{lm}(t)- \frac{1}{\tau_l}~ V_{lm}(t) 
\end{align}
with
\begin{align}
f_{l,d} &=\frac{\epsilon_d-1}{\epsilon_d+k_l} &
f_{l,0} &=\frac{\epsilon_0-1}{\epsilon_0+k_l}&
\tau_l  &=\frac{\epsilon_d+k_l}{\epsilon_0+k_l}\tau_D
\end{align}
In section \ref{sec:theory} only the dipole contribution to the reaction field, i.e., the potential contributions $V_{1m}$, is considered.

\section{EOM for the Dipole Reaction Field in Ellipsoidal cavities}
\label{App-Ell}
\label{App1}
The equation of motion for the dipole reaction field $\vec{F}$ for an ellipsoidal cavity with two of the three axis of equal length (prolate spheroid), may be derived from Eq.(\ref{eq:EOMmult}):
\begin{align}\label{eq:EOMOnsager}
 \frac{d}{dt}\vec{F}(t)=K_1 f_{1,d} \frac{d}{dt}\vec{\mu}(t)+\frac{K_1f_{1,0}}{\tau_1}\vec{\mu}(t)- \frac{1}{\tau_1} \vec{F}(t) 
\end{align}
with:
\begin{align}\label{eq:KPSp}
K_1&=\frac{3\Lambda}{bc^2} & k_1&=\frac{\Lambda}{1-\Lambda} &
f_{1,d} &=\frac{\epsilon_d-1}{\epsilon_d+k_1} &
f_{1,0} &=\frac{\epsilon_0-1}{\epsilon_0+k_1}
\end{align}
$b$ and $c$ are respectively the principal and the secondary semi-axes of the prolate spheroid, and $\Lambda$ is the prolate-spheroid depolarizing factor along the principal axis:\cite{ross1950solvent,osborn1945demagnetizing}
\begin{align}
\Lambda&=-\frac{1}{r_{bc}^2-1}\left [\frac{r_{bc}}{2\sqrt{r_{bc}^2-1}}\ln{\frac{r_{bc}+\sqrt{r_{bc}^2-1}}{r_{bc}-\sqrt{r_{bc}^2-1}}}-1\right]
\end{align}
with $r_{bc}=b/c$.
The dipole reaction field for a spherical cavity may be obtained from Eq. (\ref{eq:EOMOnsager}) in the limit of $c\rightarrow b\equiv a$ with
\begin{align}\label{eq:KSph}
K_1&=\frac{1}{a^3}         & k_1&=\frac{1}{2}
\end{align}
Eq. (\ref{eq:df_ons}) may be recovered by defining $g_0=K_1 f_{1,0}$ and $g_d=K_1 f_{1,d}$, with $\Lambda=1/3$ and $a=b=c$.

\end{document}